\documentclass[aps,pra,twocolumn]{revtex4}
\newcommand {\be}{\begin{equation}}
\newcommand {\ee}{\end{equation}}
\newcommand {\bey}{\begin{eqnarray}}
\newcommand {\eey}{\end{eqnarray}}
\usepackage{epsfig}
\usepackage{amsfonts}
\usepackage{amsmath}
\usepackage{mathbbol}

\begin{document}

\title{Communication cost of classically simulating a quantum channel with
subsequent rank-$1$ projective measurement}

\author{Alberto Montina}
\affiliation{Perimeter Institute for Theoretical Physics, 31 Caroline Street North, Waterloo, 
Ontario, Canada N2L 2Y5}

\date{\today}

\begin{abstract}
A process of preparation, transmission and subsequent projective 
measurement of a qubit can be simulated by a classical model with only
two bits of communication and some amount of shared randomness. However 
no model for $n$ qubits with a finite amount of classical communication is known at 
present. A lower bound for the communication cost can provide useful hints 
for a generalization. It is known for example that the amount of communication 
must be greater than $c\, 2^n$, where $c\simeq0.01$. The proof uses a 
quite elaborate theorem of communication complexity. Using a mathematical 
conjecture known as the ``double cap conjecture'', we strengthen this result 
by presenting a geometrical and extremely simple derivation of the lower bound 
$2^n-1$. Only rank-$1$ projective measurements are involved in the derivation.
\end{abstract}
\maketitle

One of the main differences between quantum and classical physics lies in the 
concept of state. Whereas a classical state is associated with something that 
can be observed at least in principle, a quantum state is a mathematical object 
that provides mere information about the outcome probabilities of any conceivable 
measurement. This distinction is important and makes the following question 
nontrivial: How many bits of classical communication are necessary for simulating 
the communication of $n$ qubits? Although the Hilbert space is continuous, however 
the full infinite information about the quantum state is not accessible in a 
single experimental realization. Thus, the goal of a classical simulation 
is much less than communicating the classical description of a quantum state.
Its purpose is to reproduce the measurement outcomes, performed after the 
communication, in accordance with the quantum predictions. Indeed, it was 
shown in Ref.~\cite{cerf} that the communication of one qubit and any subsequent
projective measurement can be simulated by a classical protocol using only $2.19$ 
bits of communication on average. This result was improved in Ref.~\cite{toner}, 
where Toner and Bacon reported a protocol requiring just $2$ bits of communication 
for each realization. Recently an alternative model was derived in Ref.~\cite{montina}. 
All these protocols use some resource of shared randomness, that is, the sender and 
receiver share some set of random variables.

Classical models of quantum channels are important in quantum communication
complexity~\cite{buhrman} because they can establish a limit on the advantage
that a quantum channel can provide on a classical channel for solving problems 
of distributed computing. Indeed an optimal classical model could provide a 
natural measure of the power of quantum channels. However, no generalization to 
$n$ qubits is known at present. A model for $n$ qubits can be derived from a 
protocol reported in Ref.~\cite{massar}, however it requires a two-way classical 
communication. Furthermore, the amount of communication in each execution is not 
bounded and can be arbitrarily large, although its average is finite.
Approximate models were reported in Ref.~\cite{montina} and require an 
amount of communication growing linearly with $n$. A lower bound for the 
communication cost of an exact simulation can turn out to be useful for 
finding an optimal exact protocol. First, it would lead to focus on attempts 
that satisfy the constraint. Second, the particular reasoning used for deriving 
a lower bound can suggest some general structure that the model should have, 
especially if the derivation is easily visualizable and does not require too much 
technicality. A lower bound was derived for example in Ref.~\cite{brassard}, where 
Brassard et al. showed that the amount of communication cannot be smaller that 
$c\, 2^n$, where $c\simeq0.01$. They considered the related problem of simulating
quantum entanglement with classical communication, however the result can be
easily adapted to the case of quantum channels. Their proof uses an elaborate
theorem of communication complexity, which is not easily accessible without
some technical knowledge. 

In this article, we strengthen their result by deriving the lower bound $2^n-1$. 
The derivation is extremely simple and uses a geometry conjecture known as the 
``double cap conjecture''. Although this conjecture is a mathematical open problem, 
however there are some reasons supporting its plausibility. 
Unlike in Ref.~\cite{brassard}, only rank-$1$ projective measurements are used in 
the derivation. This feature has a nontrivial consequence. Suppose that two
parties perform local two-outcome measurements on a bipartite quantum state.
In Ref.~\cite{regev} it was shown that the quantum correlation of the outcomes
can be classically simulated by using only two bits of communication for any 
dimension of the Hilbert space. However, these simulations do not correctly 
reproduce the marginal probabilities of the local outcomes. Using our result, 
we derive an exponential lower bound for the communication cost of reproducing 
the full probability distribution of the outcomes in the scenario of Ref.~\cite{regev}. 

Let us state the double cap conjecture. \newline
{\bf Conjecture 1}.
{\it Let $\cal A$ be the whole class of measurable subsets, $M$, of the 
hypersphere $S^{d-1}=\{\vec x\in{\mathbb R^d}: |\vec x|=1\}$, so that the sets
$M$ do not contain pairs of orthogonal vectors, that is,
\be\label{constr1}
\vec x_1,\vec x_2\in M \Rightarrow \vec x_1\cdot\vec x_2\ne0.
\ee
The supremum of volumes of such sets in ${\cal A}$ is equal to
the volume of two opposite caps of angular width $\pi/2$. }

\begin{figure}
\epsfig{figure=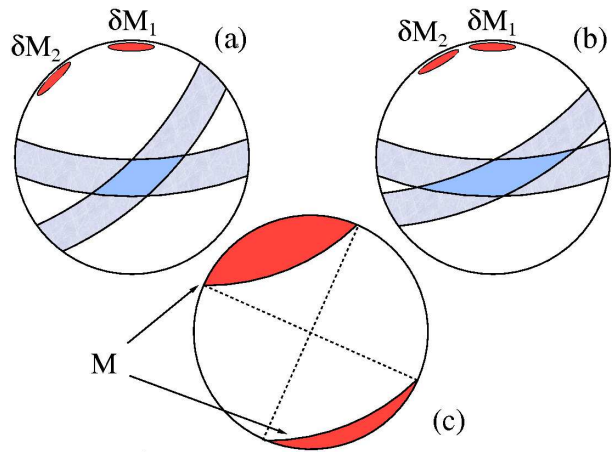,width=8.cm}
\caption{Illustrative explanation of the double cap conjecture in three
dimensions. (a) Two small sets, $\delta M_1$ and $\delta M_2$, are associated 
with two strips of forbidden points. (b) Bringing the two small sets close to 
each other, the forbidden region is made smaller. (c) The maximum set is given 
by two opposite caps with an angular width $\pi/2$.}
\label{fig1}
\end{figure}

We set the $(d-1)$-dimensional volume of the sphere equal to one.
According to conjecture 1, the maximum volume of $M$ is
\be
V_d=\frac{\int_0^{\pi/4} \sin^{d-2}x\, dx}{\int_0^{\pi/2} \sin^{d-2}x\, dx}.
\ee
For large $d$, the volume decreases exponentially as $2^{-\frac{d}{2}}\simeq 1.414^{-d}$.
Indeed, this asymptotic behaviour is supported by the 
Frankl-Wilson theorem~\cite{frankl} which gives the upper bound $1.203^{-d}$ 
for the maximum volume. A result of Raigorodskii further lowers the upper bound 
to $1.225^{-d}$~\cite{raig}, which is closer to the value given by the double cap 
conjecture. Besides these clues, it is also possible to give an intuitive reasoning 
in favour of conjecture 1. The reasoning is by construction. Suppose that we start 
with a set $M$ containing only a small region $\delta M_1$. 
This region is associated with a strip of forbidden points around a geodesic, that 
is, points that cannot be added to $M$ without breaking constraint~(\ref{constr1}). 
We can also take the specular image on the opposite 
side since this does not increase the forbidden region. Then, we add another 
small region $\delta M_2$ and its opposite image to $M$. Thus, we have to add another 
strip of forbidden points (Fig.~\ref{fig1}a). If $d$ is greater than $2$, 
it is better to make $\delta M_2$ as close as possible to $\delta M_1$, since 
this increases the overlap between the two strips and reduces the overall region 
of forbidden points (Fig.~\ref{fig1}b). In this way, we have more space for
expanding the set $M$. The procedure is repeated and other small regions are added 
close to the previous ones. This reasoning suggests that the points of the maximum 
set satisfying constraint~(\ref{constr1}) are collected
around some symmetry axis of the sphere, that is, the maximum set is the union
of two opposite caps (Fig.~\ref{fig1}c) with angular width $\pi/2$.
Notice that this reasoning does not work for $d=2$, since the forbidden
regions associated with two non-overlapping regions are always
non-overlapping. Nevertheless the double cap sets (namely two arcs)
are still maximal, although they are not the only ones.

Conjecture 1 has a natural generalization to vector spaces over the
complex field. Let $C^{2N-1}$ be the set of unit vectors in a complex
vector space $\mathbb{Z}^N$, that is, $C^{2N-1}=\{\vec x\in\mathbb{Z}^N:|\vec x|=1\}$.
There is a one-to-one correspondence between the elements in $C^{2N-1}$ and the 
points on a $(2N-1)$-hypersphere $S^{2N-1}$. Let us define the measure on 
$C^{2N-1}$ induced by the measure on $S^{2N-1}$. 

{\bf Conjecture 2.} 
{\it Let $\cal A$ be the whole class of measurable subsets, $M$, of $C^{2N-1}$ so 
that the sets $M$ do not contain pairs of orthogonal vectors. The supremum of volumes 
of such sets in $\cal A$ is equal to the volume of
a set of vectors $\vec x$ satisfying the condition $|\vec x\cdot \vec s|^2>\frac{1}{2}$, 
where $\vec s$ is some unit vector. }

Notice that the maximum set $M$ contains rays, that is, if $\vec x\in M$,
then $\alpha\vec x\in M$ for any complex number $\alpha$ of modulus $1$.
In analogy with the real case,
we concisely call the maximum set of conjecture 2 ``double cap set''.
The intuitive argument given in favour of conjecture 1 can be safely
used for supporting conjecture 2. Also in this case, the two-dimensional
case is special and the double cap set is not the only maximum
set satisfying the constraint of the conjecture.
Setting the $(2N-1)$-dimensional volume of the hypersphere equal to $1$,
the volume of the double caps in the complex case is 
\be\label{complex_vol}
U_N=\frac{\int_0^{\pi/4}\cos x\,\sin^{2N-3} x}{\int_0^{\pi/2}\cos x\,\sin^{2N-3} x}
=2^{1-N}.
\ee
The derivation of Eq.~(\ref{complex_vol}) is as follows. Let us denote 
by $\vec y\in {\mathbb{R}^{2N}}$ the 
$2N$-dimensional real vector associated with a unit complex vector $\vec x$. 
In the real notation, the `complex' double cap is given by any vector
$\vec y$ such that 
\be\label{ineq}
(\vec y\cdot\vec s_1)^2+(\vec y\cdot\vec s_2)^2>\frac{1}{2},
\ee
where $\vec s_1$ and $\vec s_2$ are two suitable orthogonal unit vectors.
The vector $\vec y$ can be written in the form
\be
\vec y=\cos\theta \vec u_1+\sin\theta \vec u_2,
\ee
where $\vec u_1$ is a unit vector of the two-dimensional subspace spanned by 
$\vec s_1$ and $\vec s_2$, whereas $\vec u_2$ is a vector in the $(2N-2)$-dimensional
orthogonal complement. By inequality~(\ref{ineq}) we have that 
the double caps are defined by the inequalities $0\le\theta<\pi/4$
and $3\pi/4<\theta\le\pi$. Thus, it is easy to realize that the double cap volume is 
\be
U_N=\frac{2}{W_{2N-1}}\int_0^{\pi/4} d\theta w_1(\cos\theta)w_{2N-3}(\sin\theta),
\ee
where $w_d(r)$ is the volume of a $d$-dimensional hypersphere of radius $r$
and $W_d\equiv w_d(1)$. This equation gives Eq.~(\ref{complex_vol}).

With these premises, let us consider the following scenario of
quantum communication. Suppose that there are two parties, Alice and
Bob. Alice prepares $n$ qubits in a quantum state $|\psi\rangle$,
then she sends them to Bob, who finally performs a rank-$1$ projective
measurement of the qubits. Let us denote by $|\phi\rangle\langle\phi|$
the measured observable. Suppose now that Alice and Bob want
to simulate this scenario using a classical channel. How
many bits of communication are required by the simulation?
For $n=1$, the Toner-Bacon model shows that $2$ bits of communication
are sufficient. Using the double cap conjecture we will derive the
lower bound $-\log_2 V_N$, where $N\equiv2^n$ is the Hilbert
space dimension. This result will be improved using the generalized
conjecture 2, which gives the lower bound $2^n-1$.

Besides the communication resource in the classical simulation,
the two parties are also allowed to share 
some common random variable $X$.  In other words, 
before the game begins, Alice and Bob 
receive an identical list of random values of $X$ with probability distribution
$\rho(X)$. The variable $X$ could be a real number, a vector or a set of vectors. 
No constraint on this shared resource is given. The classical protocol
is as follows. Alice has a classical description of the state $|\psi\rangle$ and 
generates an index $k$ with probability distribution $\rho(k|X,\psi)$. The
index $k$ takes $R$ possible values. Then, she sends $k$ to Bob.
This requires $\log_2 R$ bits of communication. Finally, Bob generates
an event $|\phi\rangle$ with probability $P(\phi|k,X)$. The protocol
simulates the quantum channel and the subsequent measurement if
\be\label{equiv}
\sum_k\int dX P(\phi|k,X)\rho(k|X,\psi)\rho(X)=|\langle\phi|\psi\rangle|^2.
\ee
The probability functions satisfy the constraints
\be\label{constr}
\begin{array}{c}
0\le P(\phi|k,X)\le1, 
\vspace{1mm} \\
\sum_k \rho(k|X,\psi)=1, \hspace{3mm}
\rho(k|X,\psi)\ge0,
\vspace{1mm} \\
\int dX\rho(X)=1,
\hspace{3mm} \rho(X)\ge0.
\end{array}
\ee

This protocol has to satisfy the following general property.
Let $\Omega_k(X)$ be the set of vectors $|\psi\rangle$ such that
$\rho(k|X,\psi)$ is different from zero. \newline
{\bf Lemma 1.} For every value of $k$ and $X$, $\Omega_k(X)$ does not
contain any pair of orthogonal vectors. 

A similar lemma was used for example in Refs.~\cite{montina2,montina3},
where we proved that, in a Markov hidden variable theory, the number
of continuous variables describing $n$ qubits grows exponentially
with $n$.
More precisely, we should say that the property stated by Lemma 1 
holds apart from a zero-probability subset of values of $X$. 
The physical meaning of lemma 1 is clear. All Bob knows about the 
state $|\psi\rangle$ is contained in the values of the index $k$ 
and $X$. Given these values, Bob knows that $|\psi\rangle$
is in a subset $\Omega_k(X)$. But if $\Omega_k(X)$ contains
two orthogonal vectors, $|\psi_1\rangle$ and $|\psi_{-1}\rangle$,
and he wishes to measure the observable $|\psi_1\rangle\langle\psi_1|$,
he has no way to produce an outcome that is compatible with both the 
distinct states $|\psi_1\rangle$ and $|\psi_{-1}\rangle$. Thus,
$\Omega_k(X)$ cannot contain pairs of orthogonal vectors. Here
the formal proof. \newline
{\bf Proof by contradiction}. Suppose that there is a value 
$l$ of $k$, for some $X$, such that $\Omega_{l}(X)$ contains two 
orthogonal vectors, $|\psi_1\rangle$ and $|\psi_{-1}\rangle$. Thus,
\be
\rho(l|X,\psi_n)\ne0, \; \text{ for } n=\pm1.
\ee
From Eq.~(\ref{equiv}), we have that
\be
\sum_k\int dX P(\psi_n|k,X)\rho(k|X,\psi_n)\rho(X)=1.
\ee
Since $\rho(l|X,\psi_n)\ne0$ for some $X$ and the probability functions
satisfy constraints (\ref{constr}), it is easy to realize
that 
\be\label{eq_P}
P(\psi_n|l,X)=1 \text{  for  } n=\pm1.
\ee
Similarly, we have that
\be
\sum_k\int dX P(\psi_{-n}|k,X)\rho(k|X,\psi_n)\rho(X)=0,
\ee
which implies that
\be
P(\psi_{-n}|l,X)=0 \text{  for  } n=\pm1,
\ee
but this equation is in contradiction with Eq.~(\ref{eq_P}).
The lemma is proved. $\square$

Since the classical model has to work for any $|\psi\rangle$, we have 
that the union $\cup_k \Omega_k(X)$ contains every vector of the
Hilbert space for any $X$. Thus, if ${\cal V}[\Omega_k(X)]$ is
the volume of the set $\Omega_k(X)$, then 
\be\label{sum_set}
\sum_k{\cal V}[\Omega_k(X)]\ge1.
\ee
Notice that this equation and lemma~1 hold for both real and
complex Hilbert spaces.

At this point, let us state the main theorem. \newline
{\bf Theorem 1.} 
{\it If conjecture $1$ is true, then a process of
preparation, transmission and subsequent rank-$1$ projective
measurement of $n$ qubits cannot be simulated with an
amount of communication smaller than $-\log_2 V_{N}$, where
$N\equiv 2^n$. If conjecture $2$ is also correct, then
the lower bound is increased to $2^n-1$. }

{\bf Proof.} The proof is trivial. Using conjecture $1$, lemma~1 (adapted
to the case of a real Hilbert space) and Eq. (\ref{sum_set}), we have that
\be
R V_{N}\ge 1,
\ee
where $R$ is the number of values that the index $k$ can take.
Thus, the minimal number of bits is $\log_2 R=-\log_2 V_{N}$. Similarly,
by conjecture $2$ we have that the lower bound is $-\log_2 U_N=2^n-1$.
$\square$

Notice that the lower bounds are not just on the average number
of bits, but on the minimal number of bits communicated in
each single execution of the simulation. 
Apart from $N=2$, the volume $V_N$ is always strictly larger than 
$U_N$, thus the lower bound $2^n-1$ is stronger than $-\log_2 V_{N}$.
For large $N$, $V_N$ is well-approximated by the formula
\be
V_N\simeq \frac{2^{-\frac{N}{2}+2}}{\sqrt{2\pi N}}.
\ee

Even if the double cap conjecture was false, it is possible to prove a 
slightly weaker theorem. A result in Ref.~\cite{raig} implies that the 
maximum volume on a hypersphere under constraint~(\ref{constr1}) must be
smaller than $(\theta+\epsilon)^{-N}$ for each $\epsilon>0$ and all sufficiently
large $N$, with $\theta\equiv(2/\sqrt{3})^{\sqrt2}\simeq1.225$. Thus,
using the same proof of theorem~1, we have the following.\newline
{\bf Theorem 2.} 
{\it A process of preparation, transmission and subsequent rank-$1$ projective
measurement of $n$ qubits cannot be simulated with an amount of communication 
smaller than $2^n \log_2 (\theta+\epsilon)$ for each $\epsilon>0$ and
all sufficiently large $n$. } \newline
This theorem establishes the asymptotic lower bound $0.293\times 2^n$
for the communication cost.

Taking for granted the intuitive double cap conjecture and its `complex'
generalization, the proved theorem 1 is extremely simple and has a
geometric interpretation. Apart from this advantage, it strengthens 
the result in Ref.~\cite{brassard} in two ways. First, it gives a stronger 
lower bound for the communication cost. Second, the derivation uses only 
rank-$1$ projective measurements. This last feature has a nontrivial 
consequence. Let us consider the scenario discussed in Ref.~\cite{regev}.
Two parties, Alice and Bob, share a bipartite quantum state and perform
local two-outcome measurements. Each local outcome, $s_a$ and $s_b$, is a bit 
taking values $\pm1$. In Ref.~\cite{regev} it was shown that the correlation 
$\langle s_a s_b\rangle$ can be reproduced by a classical simulation with only 
$2$ bits of communication regardless of the dimension of the Hilbert space.
However, the reported models does not reproduce the correct marginal
distributions of $s_a$ and $s_b$. Indeed, Theorem~1 
(but the weaker theorem~2 would be sufficient) implies
that an exact reproduction of the full probability distribution
requires an exponentially growing amount of communication. 
It is known that any classical protocol that
simulates $n$ ebits can be converted into a classical protocol
simulating a quantum channel of $n$ qubits with a negligible increase of
communication. A general conversion method is reported for example
in Ref.~\cite{montina}. The additional amount of communication
is equal to the number of ebits on average. In particular, the conversion 
of a model in the scenario of Ref.~\cite{regev} gives a classical
model of a quantum channel with a subsequent two-outcome measurement.
Since theorem~1 holds for rank-$1$ projective measurements
and these measurements are a subset of the class of the
two-outcome measurements, we have automatically the following.\newline
{\bf Corollary.} {\it The minimal amount of communication needed for
simulating 
two-outcome measurements on maximally entangled bipartite
quantum states grows exponentially with the number of ebit.}

In particular, theorem~1 gives $2^n-1-n$ as the lower bound of the average amount 
of communication. Indeed, suppose that there is a model of entanglement for
two-outcome measurements and this model requires less than $2^n-1-n$ bits of communication 
on average. Then it is possible to convert it into a model of quantum channel 
with a subsequent rank-$1$ measurement. But this model would require less than 
$2^n-1$ bits of communication, in contradiction with theorem~1. A recent 
result in Ref.~\cite{boaz} gives the asymptotic weaker lower bound $2^{n/3}$, 
which is derived in a scenario where traceless two-outcome measurements are
considered.

In conclusion, using a plausible mathematical conjecture, we have derived 
the lower bound $2^n-1$ for the classical communication cost of simulating 
the quantum communication of $n$ qubits with subsequent rank-$1$ projective 
measurement. Our proof is simple and can provide useful hints for finding
optimal one-way protocols that simulate quantum channels.

{\it Acknowledgements.} I wish to thank Ben Toner for useful discussions
and Guillaume Aubrun for bringing to my knowledge that the double cap conjecture
is an open mathematical problem.
Research at Perimeter Institute for Theoretical Physics is
supported in part by the Government of Canada through NSERC
and by the Province of Ontario through MRI.

\end{document}